\documentclass[aps,prd,amssymb,nofootinbib,twocolumn,epsf,floatfix]
{revtex4}
\usepackage[usenames]{color}
\usepackage[normalem]{ulem} 
\usepackage{graphicx} 
\usepackage{subfigure}  
\usepackage{amssymb}  
\usepackage{amsmath}
\usepackage{mathrsfs}
\usepackage{epstopdf}
\usepackage{float}
\usepackage{booktabs} % for tabular with \toprule, \midrule, \bottomrule
\usepackage{microtype}
\usepackage[colorlinks,breaklinks]{hyperref}
\begin{document}
 
%%%%%%%%%%%%%%%%%%%%%%%%%%%%%%%%%%%%%%%%%%%%%%%%%%%%%%%%%%%%%%%%%%%%%%%%%%%%%%%
% Solving Einstein's Equations on Non-Orientable Manifolds
%%%%%%%%%%%%%%%%%%%%%%%%%%%%%%%%%%%%%%%%%%%%%%%%%%%%%%%%%%%%%%%%%%%%%%%%%%%%%%%

\title{Solving Einstein's Equation Numerically\\ on Manifolds with
Non-Orientable Spatial Slices}

\author{Fan Zhang\,${}^{a,b}$ and Lee Lindblom\,${}^c$}

\affiliation{${}^a$Institute for Frontiers in Astronomy and Astrophysics,
  Beijing Normal University, Beijing 102206, China}

\affiliation{${}^b$Gravitational Wave and Cosmology Laboratory,
  School of Physics and Astronomy,
  Beijing Normal University, Beijing 100875, China}

\affiliation{${}^c$ Department of Physics, University 
  of California at San Diego, La Jolla, CA 92093, USA}

\date{\today}

\begin{abstract}
  This paper presents solutions to Einstein's equation---and the
  numerical methods used to construct them---that describe simple
  cosmological models on manifolds with compact non-orientable spatial
  slices.  These solutions have been constructed on a selection of
  manifolds having positive, negative, and vanishing spatial scalar
  curvatures.  One example is shown to be indistinguishable locally
  from a homogeneous Friedman cosmological model, others are
  constructed with significant inhomogeneities.  Together these
  examples are used to explore the strengths and the limitations of
  the numerical methods used in this study, and to test the code used
  to implement them.
\end{abstract}

\maketitle

%%%%%%%%%%%%%%%%%%%%%%%%%%%%%%%%%%%%%%%%%%%%%%%%%%%%%%%%%%%%%%%%%%%%%%%%%%%%%%%
% Introduction
%%%%%%%%%%%%%%%%%%%%%%%%%%%%%%%%%%%%%%%%%%%%%%%%%%%%%%%%%%%%%%%%%%%%%%%%%%%%%%%
\section{Introduction}
\label{s:Introduction}
%%%%%%%%%%%%%%%%%%%%%%%%%%%%%%%%%%%%%%%%%%%%%%%%%%%%%%%%%%%%%%%%%%%%%%%%%%%%%%%

Einstein's equation determines the geometry of space-time given
initial data for the matter fields (and any gravitational waves) that
may be present on a spacelike surface.  This equation does not,
however, determine the topology of that initial spatial slice.  That
topology must be specified independently to insure the solutions
faithfully model a particular physical problem.  Solutions to
Einstein's equation have been studied on manifolds with a small number
of simple topologies, e.g., with initial surfaces having $S^3$, $T^3$,
or $S^2 \times S^1$ topologies, etc.  But little is known about
solutions on manifolds from most of the infinite collection of
possible topologies.  Current observations do not determine or even
constrain the possible topology of our universe.  Therefore it seems
reasonable to explore the properties of solutions on manifolds having
a wide variety of possible topologies.  To that end robust numerical
methods have been developed---and implemented in an efficient computer
code---that facilitate the study of solutions to Einstein's equations
on manifolds with arbitrary spatial
topologies~\cite{Lindblom2013,Lindblom2014,Lindblom2015,Zhang2022}.
This study extends that work by constructing and analyzing solutions
to Einstein's equations on manifolds with non-orientable spatial
slices.

Space-times with non-orientable topologies are often dismissed as
nonphysical because analysis shows that only certain types of global
spinor structures are allowed on such
manifolds~\cite{Geroch1968,Geroch1970,Grinstein1987}.  These arguments
probably rule out the possibility that the observable part of our
universe is time or space non-orientable.  However, they do not rule
out the possibility that spatial slices could become non-orientable
when extended beyond our cosmic horizon.

This study has three primary goals.  The first is to demonstrate that
it is possible to find solutions to Einstein's equation on manifolds
with non-orientable topologies.  The second is to explore whether
homogeneous solutions on non-orientable manifolds can be distinguished
in some way locally from the standard homogeneous Friedman cosmological
models.  The third goal is to explore and evaluate the strengths and
limitations of the numerical methods that have been developed to solve
Einstein's equation on manifolds with arbitrary spatial topologies.
For example, the code developed to solve the covariant representation
of the Einstein evolution equations (needed for non-trivial
topologies) has only been tested previously by evolving solutions with
small amplitude gravitational wave perturbations on a manifold with
$S^3$ spatial topology~\cite{Lindblom2014}.  In this study the code is
tested more robustly by evolving the fully nonlinear equations long
enough to expand the spatial volumes of the solutions by a factor of
ten.

The remainder of this paper is organized as follows.  The simple
non-orientable manifolds used in this study are described in
Sec.~\ref{s:SimpleManifolds}.  This section describes the multi-cube
representations of these manifolds, including the construction of the
reference metrics needed to define their differentiable structures.
This section also demonstrates that the manifolds included in this
study satisfy the topological conditions needed to admit global Dirac
and Majorana (but not Weyl) spinor structures.

Section~\ref{s:InitialData} describes the methods used to construct
simple initial data that satisfy the Einstein constraint equations on
these manifolds.  These initial data were found by solving a version
of the Einstein constraints for a spacetime devoid of matter except
for a cosmological constant.  The value of this constant is chosen in
each example to produce a geometry with a homogeneous spatial Ricci
scalar curvature.  The quality of these numerical solutions are
evaluated, and the convergence of the numerical methods used to solve
the elliptic constraint equations is illustrated.

Section~\ref{s:NonOrientableCosmologies} describes and analyzes the
numerical evolution of these initial data which produce simple
non-orientable cosmological models.  One of these models is shown to
be indistinguishable locally from one of the standard homogeneous
Friedman cosmological models.  The other examples have significant
inhomogeneities that grow rapidly from inhomogeneities present in the
initial data for these cases.  Together these examples provide an
opportunity to evaluate the strengths and limitations of the numerical
methods and their implementation in the computer code used to solve
the hyperbolic Einstein evolution equations.  The covariant symmetric
hyperbolic representation of the Einstein evolution equations used in
this study is described briefly in
Appendix~\ref{s:SymmetricHyperbolicSystem}.
Section~\ref{s:Discussion} presents a summary and discussion of the
results of this study, including ideas that could be pursued to
address some open questions.

%%%%%%%%%%%%%%%%%%%%%%%%%%%%%%%%%%%%%%%%%%%%%%%%%%%%%%%%%%%%%%%%%%%%%%%%%%%%%%%
% Simple Non-orientable Manifolds
%%%%%%%%%%%%%%%%%%%%%%%%%%%%%%%%%%%%%%%%%%%%%%%%%%%%%%%%%%%%%%%%%%%%%%%%%%%%%%%
\section{Simple Non-Orientable Manifolds}
\label{s:SimpleManifolds}
%%%%%%%%%%%%%%%%%%%%%%%%%%%%%%%%%%%%%%%%%%%%%%%%%%%%%%%%%%%%%%%%%%%%%%%%%%%%%%%

This section describes the simple non-orientable manifolds on which
solutions to Einstein's equation have been constructed for this
study. Each of these four-dimensional manifolds has the topology of a
globally hyperbolic spacetime, i.e. its topology has the form
$\mathbb{R}\times\Sigma$, where $\Sigma$ is a compact non-orientable
three-dimensional manifold without boundary.  This study
uses multi-cube representations of the three-manifolds, $\Sigma$, to
facilitate the numerical analysis.  These multi-cube representations
(introduced in Ref.\cite{Lindblom2013}) are collections of
non-overlapping cubic regions which serve as coordinate patches that
cover the manifold. 

\begin{figure}[!h]  
\centering
\subfigure{
  \includegraphics[height=0.13\textwidth]{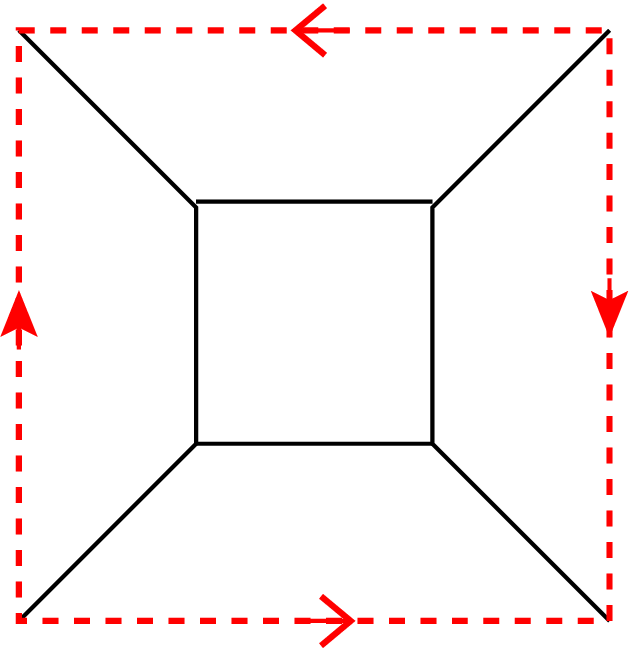}
}\hspace{0.15cm}
\subfigure{
  \includegraphics[height=0.13\textwidth]{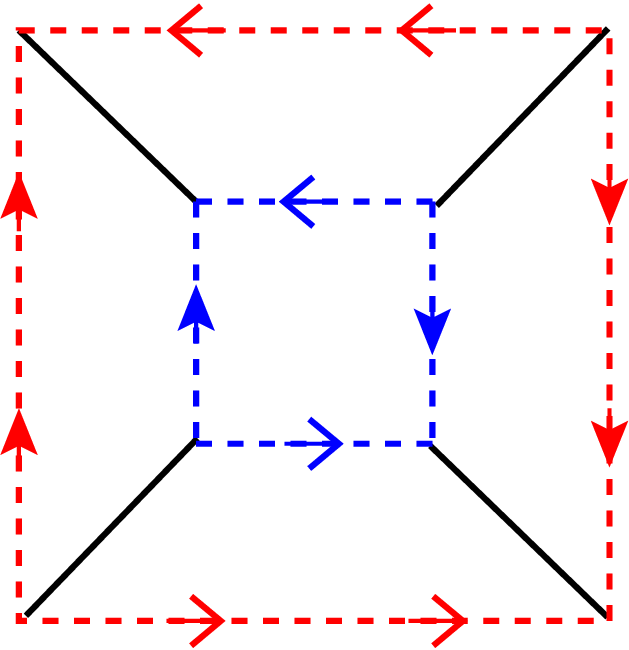}
}\hspace{0.15cm}
\subfigure{
  \includegraphics[height=0.13\textwidth]{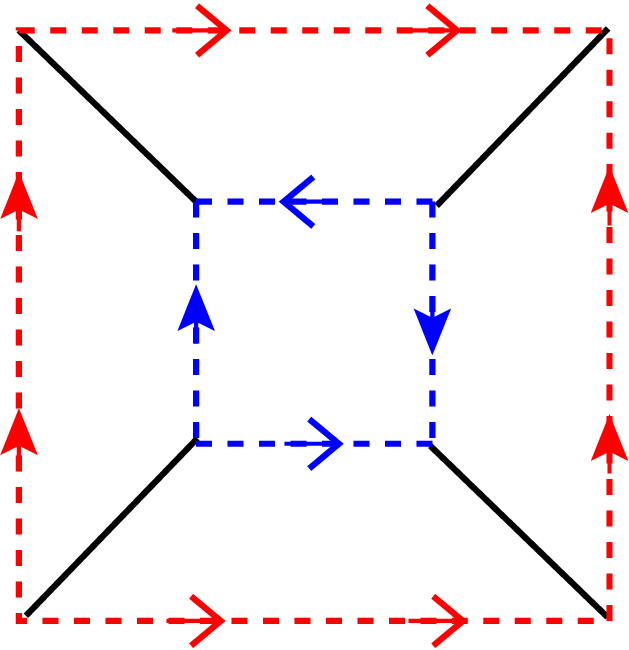}
}
\caption{\label{f:NonOrientableExamples} This figure illustrates the
  $+z$ faces of the multi-cube blocks 
  compact non-orientable three-manifolds used in this study: (from
  left to right) $P^2\times S^1$, $P^2\# P^2\times S^1$, and $P^2\#
  T^2 \times S^1$.  The solid (black) lines represent the $x$ and $y$
  boundary faces between adjacent blocks, while the dashed lines
  represent faces that are identified with the faces having similar
  arrows, in the directions indicated by those arrows.  The $+z$ face
  of each block is identified with the $-z$ face of that block.}
\end{figure}

Figure~\ref{f:NonOrientableExamples} illustrates the multi-cube
structures for three of the manifolds used in this study: $P^2\times
S^1$, $P^2\#P^2\times S^1$ and $P^2\# T^2\times S^1$.  Each of these
manifolds is the Cartesian product of a non-orientable compact
two-manifold with the circle, $S^1$. $P^2$ represents the
two-dimensional real projective plane and $T^2$ the two-torus, while
$\times$ indicates a Cartesian product and $\#$ a
connected sum.\footnote{The connected sum,
  $M_1\# M_2$, of two $m$-dimensional manifolds is formed by deleting
  an open ball from each manifold and gluing together the resulting
  boundary spheres.} These manifolds are represented as a collection
of non-overlapping cubic regions whose faces are identified in the
prescribed way.

Figure~\ref{f:NonOrientableExamples} illustrates the $+z$ faces of the
cubic regions representing these manifolds, including the way the $x$
and $y$ faces are identified.  The $+z$ and $-z$ face of each of these
cubic region is identified to implement the Cartesian product with
$S^1$.  For example, the figure on the left in
Fig.~\ref{f:NonOrientableExamples} illustrate the $+z$ faces of the
five cubic regions used to represent $P^2\times S^1$.  The solid
(black) lines in this figure represent the identifications of the $x$
and $y$ faces of the boundaries between neighboring cubes (which are
distorted in this figure to illustrate the connections between them),
while the oriented dashed (red or blue) lines represent the
identifications of the outer faces in this figure. The center and
right figures illustrate the four cubic regions used to represent
$P^2\# P^2\times S^1$ and $P^2\# T^2\times S^1$ respectively.

\begin{figure}[!h]
  \centering
  \includegraphics[height=0.2\textwidth]{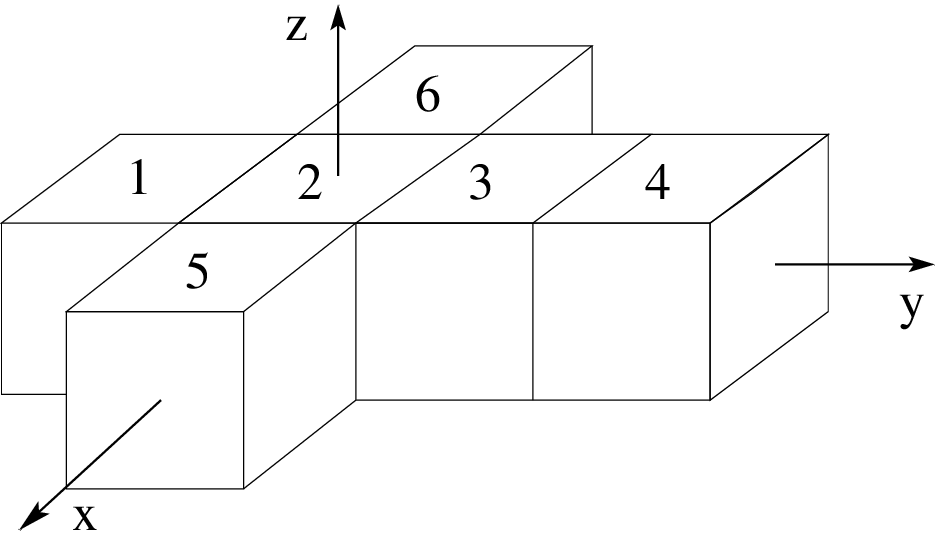}
  \caption{\label{f:S2xtildeS1} This figure illustrates the multi-cube
    structure used in this study for the compact non-orientable
    three-manifold $S^2 \tilde\times S^1$. }
\end{figure}

Figure~\ref{f:S2xtildeS1} illustrates the other compact non-orientable
three-manifold used in this study: $S^2\tilde\times S^1$, where $S^2$
and $S^1$ are the two-sphere and the circle respectively, while
$\tilde\times$ represents a twisted Cartesian product (described
below).  The locations of the blocks in this structure are the same as
those used in the multi-cube structure of $S^2\times S^1$ introduced
in Ref.~\cite{Lindblom2013}.\footnote{This representation
  of $S^2\times S^1$ is analogous to the ``cubed sphere''
  representation of $S^2$ introduced in Refs.~\cite{Sadourny1972,
    Ronchi1996}, and used in the representation of $S^2\times R$ in
  Ref.~\cite{Lehner2005}.} In $S^2\times S^1$ the $+z$ face of each
block is identified with the $-z$ face of that block.  In $S^2
\tilde\times S^1$ the $+z$ face of each block is identified with the
$-z$ face of the block containing its polar opposite points in $S^2$.
In this way the $+z$ face of region $\mathcal{B}_1$ is identified with
the $-z$ face of region $\mathcal{B}_3$, the $+z$ face of region
$\mathcal{B}_2$ is identified with the $-z$ face of region
$\mathcal{B}_4$, the $+z$ face of region $\mathcal{B}_5$ is identified
with the $-z$ face of region $\mathcal{B}_6$, etc.

The Euler characteristic of the Cartesian product of two manifolds is
the product of the Euler characteristics of the individual manifolds.
Since the Euler characteristic of $S^1$ is zero, the Euler
characteristics of the manifolds $P^2\times S^1$, $P^2\# P^2\times
S^1$, and $P^2\# T^2 \times S^1$ are also zero. The multi-cube
structure of the manifold $S^2 \tilde\times S^1$ shown in
Fig.~\ref{f:S2xtildeS1} has the same number of vertices, edges, faces,
and volumes as the multi-cube structure of $S^2 \times S^1$, therefore
its Euler characteristic is also zero.  Dirac and Majorana spinor
structures do exist on a non-orientable manifold if the Euler
characteristic of that manifold is an even
integer~\cite{Grinstein1987}.  It follows that all the non-orientable
manifolds included in this study do admit global Dirac and Majorana
spinor structures. These manifolds do not admit global Weyl spinor
structures~\cite{Geroch1968,Geroch1970}, however, Weyl spinor
structures do exist on any orientable submanifold.  These orientable
submanolds can extend in some cases through the cosmic horizon.  Therefore
it seems unlikely that non-orietability on a global scale can be ruled
out by observations at this time.

The Cartesian coordinates within each cubic region in a multi-cube
representation of a manifold serve as local coordinate patches on
which scalar, vector and tensor fields can be represented on this
manifold.  The maps used to identify the faces of neighboring cubes in
these structures are used to determine when scalar fields are
continuous across those faces.  Since the cubic regions do not
overlap, however, in general these boundary maps do not determine when
vector and tensor fields are continuous across those interfaces.  More
geometric structure must be provided to determine the continuity of
vector and tensor fields across the interface boundaries between cubic
regions.

The additional structure needed to determine the continuity and
differentiability of vector and tensor fields across the interface
boundaries in a multi-cube representation can be provided by a
``reference metric'' that is suitably smooth (typically at least
$\mathcal{C}^1$)~\cite{Lindblom2013}.  No ``trivial'' reference metric
is known for most manfolds, however, suitable reference metrics can be
constructed numerically using the method described in detail in
Ref.~\cite{Lindblom2015} for two-dimensional multi-cube structures and
in Ref.~\cite{Lindblom2022} for three-dimensional structures.  This
method constructs a sequence of increasingly smooth metrics that
ensure the final reference metric, $\tilde g_{ij}$, is at least
$\mathcal{C}^{2-}$ across each cube face and along each cube
edge.\footnote{The construction of a suitable reference metric begins
by constructing a flat metric in the neighborhood of each vertex point
in the multi-cube structure.  These flat metrics are constructed to
ensure there are no conical singularities along any of the cube edges.
These local flat metrics are then combined using a partition of unity
to form a global $\mathcal{C}^0$ metric.  An additional step makes
these metrics $\mathcal{C}^1$ on the faces between cubes, by solving a
fourth-order elliptic equation that adjusts the normal derivatives of
the metrics at the cube faces in a way that makes the extrinsic
curvatures continuous between neighboring cubes.  The continuity of
the extrinsic curvatures ensures the metrics are $\mathcal{C}^1$
across the cube faces.  Details of these steps are given in
Ref.\cite{Lindblom2022}.}

The method of constructing the reference metric, $\tilde g_{ij}$, on
multi-cube structures depends on the details of how each cubic region
is connected to each of its neighbors: across each cube face, and
along each cube edge.  However this method does not depend on the
global topological properties of the manifold, such as its
orientability.  This method can therefore be used to construct
reference metrics on any non-orientable three-manifold, and has been
used successfully to construct $\mathcal{C}^1$ metrics, $\tilde
g_{ij}$, on each of the non-orientable manifolds included in this
study: $P^2\times S^1$, $P^2\#P^2\times S^1$, $P^2\# T^2\times S^1$,
and $S^2\tilde\times S^1$.  The resulting positive-definite metrics
$\tilde g_{ij}$ on these manifolds are smooth within each cubic
coordinate chart, and are continuous and differentiable in the
appropriate senses across the interfaces between cubic regions.  These
reference metrics, $\tilde g_{ij}$, are then used to construct the
Jacobians that define the differentiable structure that determines the
global continuity and differentiability of vectors and tensors on
these manifolds~\cite{Lindblom2015,Lindblom2022}.

%%%%%%%%%%%%%%%%%%%%%%%%%%%%%%%%%%%%%%%%%%%%%%%%%%%%%%%%%%%%%%%%%%%%%%%%%%%%%%%
% Simple Initial Data
%%%%%%%%%%%%%%%%%%%%%%%%%%%%%%%%%%%%%%%%%%%%%%%%%%%%%%%%%%%%%%%%%%%%%%%%%%%%%%%
\section{Simple Initial Data}
\label{s:InitialData}
%%%%%%%%%%%%%%%%%%%%%%%%%%%%%%%%%%%%%%%%%%%%%%%%%%%%%%%%%%%%%%%%%%%%%%%%%%%%%%%

This section describes the methods used in this study to find
solutions to the Einstein initial value constraint equations. The goal
is to construct initial data that can be used to evolve simple
homogeneous cosmological models on these non-orientable manifolds.
The Einstein constraint equations limit the allowed initial data for
the spatial metric $g_{ij}$ and extrinsic curvature $K_{ij}$ on a
spacelike hypersurface.  The simplest solutions to these equations are
determined by specifying a conformal metric, $\tilde g_{ij}$, and a
constant, $K$, that functions as the trace of the extrinsic curvature
on this initial surface.  The physical metric $g_{ij}$ and extrinsic
curvature $K_{ij}$ are given by
\begin{eqnarray}
g_{ij} &=& \phi^4 \tilde g_{ij},
  \label{e:gmetricDef}\\
  K_{ij} &=& \tfrac{1}{3}\phi^4 \tilde g_{ij} K,
  \label{e:KDef}
\end{eqnarray}
where $\phi$ is a conformal factor.  This conformal factor is
determined by solving the Einstein constraint equations, which for
this simple case reduce to a second-order elliptic partial
differential equation for $\phi$.

This study follows the methods described in Ref.~\cite{Zhang2022} for
solving the Einstein constraint equations numerically on manifolds
with non-trivial topologies. The conformal metrics $\tilde g_{ij}$
used in this study are the reference metrics constructed in
Sec.~\ref{s:SimpleManifolds} for each manifold.  This study also
assumes the average matter content in these solutions is negligible
compared to the cosmological constant $\Lambda$.  With these
assumptions the Einstein constraint equation reduces to
\begin{eqnarray}
  \tilde\nabla^i\tilde\nabla_i\phi &=& \textstyle\frac{1}{8}\phi\,
  \tilde R +\textstyle\frac{1}{12}\phi^5\, (K^2-3\Lambda),
  \label{e:CMCConstraintS}
\end{eqnarray}
where $\tilde \nabla_i$ and $\tilde R$ are the covariant derivative
and Ricci scalar curvature respectively associated with the conformal
metric $\tilde g_{ij}$.

The integral of the left side of Eq.~(\ref{e:CMCConstraintS}) vanishes
on any compact manifold. Therefore the constants $K$ and $\Lambda$
must be chosen in a way that makes it possible for the integral of the
right side of this equation to vanish as well.  Convenient choices for
these constants would produce solutions to
Eq.~(\ref{e:CMCConstraintS}) with $\phi\approx 1$.  Such choices can
be identified by setting $\phi=1$ in the expression on the right side
of Eq.~(\ref{e:CMCConstraintS}) and integrating over the manifold.
Setting this integral to zero results in the values,
\begin{equation}
  K^2 - 3\Lambda = - \textstyle\frac{3}{2}\langle \tilde R\, \rangle,
  \label{e:CMCParameterChoices}
\end{equation}
where $\langle \tilde R\, \rangle$ is the average value of the
conformal scalar curvature $\tilde R$,
\begin{equation}
  \langle \tilde R\, \rangle
  = \frac{\int \sqrt{\tilde g}\,\tilde R\, d^{\,3}x}
  {\int \sqrt{\tilde g}\, d^{\,3}x},
  \label{e:AvarageRtildeDef}
\end{equation}
where $\tilde g$ represents the determinant of the conformal metric
$\tilde g_{ij}$.  This choice transforms Eq.~(\ref{e:CMCConstraintS})
into the form 
\begin{eqnarray}
  \tilde\nabla^i\tilde\nabla_i\phi &=& \textstyle\frac{1}{8}\phi\,
  \left(\tilde R - \phi^4 \langle \tilde R\,\rangle\right).
  \label{e:CMCConstraintSAlt}
\end{eqnarray}
This equation has the exact solution $\phi=1$ in the constant scalar
curvature case $\tilde R =\langle \tilde R\,\rangle$, and admits
solutions on all the manifolds studied here. The integral of the right
side of Eq.~(\ref{e:CMCConstraintS}) must vanish for any solution
$\phi$.  If $\phi>0$ and $\tilde R>0$ this integral can vanish only if
$K^2 - 3\Lambda<0$.  Thus no $\phi>0$ solution can exist to
Eq.~(\ref{e:CMCConstraintS}) when $\tilde R>0$ unless the cosmological
constant satisfies the inequality, $\Lambda > \tfrac{1}{3}K^2 \geq 0$.

The Ricci scalar curvature $R$ determined from the physical metric
$g_{ij}$ is related to the Ricci scalar curvature $\tilde R$
associated with the conformal metric $\tilde g_{ij}$ by the expression
\begin{eqnarray}
R &=& \phi^{-4}\, \tilde R - 8 \phi^{-5}\, \tilde \nabla^i\tilde\nabla_i\phi.
  \label{e:ConformalRicciScalar}
\end{eqnarray}
Solutions to the Einstein constraint Eq.~(\ref{e:CMCConstraintSAlt}) therefore
have physical Ricci scalar curvatures given by
\begin{eqnarray}
  R&=&\langle\tilde R\,\rangle.
  \label{e:ConstantRCondition}
\end{eqnarray}
These simple solutions to the Einstein constraint equations therefore
have the property that the physical scalar curvature $R$ is constant,
$R=\langle\tilde R\,\rangle$.  Thus the conformal factor $\phi$ is the
solution to the Yamabe problem~\cite{Yamabe1960} that transforms
$\tilde g_{ij}$ into the constant scalar curvature metric $g_{ij}$.

For this study the differential Eq.~(\ref{e:CMCConstraintSAlt}) has
been solved numerically using the pseudo-spectral methods implemented
in the \verb!SpEC! numerical relativity code (developed initially by
the Caltech and Cornell numerical relativity groups) as described in
some detail in Refs.~\cite{Pfeiffer2003,Zhang2022}.  These methods
solve the non-linear Eq.~(\ref{e:CMCConstraintSAlt}) by minimizing the
discrete version of the residual $\mathcal{E}$ defined by
\begin{equation}
  \mathcal{E}=\tilde\nabla^i\tilde\nabla_i\phi 
  -\textstyle\frac{1}{8}\phi\, \left(\tilde R
  - \phi^4 \langle \tilde R\,\rangle\right).
  \label{e:CMCResidualDef}
\end{equation}
Table~\ref{t:TableI} lists $\langle \tilde R\,\rangle$ defined in
Eq.~(\ref{e:AvarageRtildeDef}) and the physical volumes
$\mathcal{V}$ defined by
\begin{equation}
  \mathcal{V} = \int \sqrt{g}\, d^{\,3}x,
  \label{e:PhysicalVolumeV}
\end{equation}
(where $g$ is the determinant of the physical metric) for the
geometries constructed for each of the manifolds included in this
study.  These volumes $\mathcal{V}$ measure the physical ``sizes'' of
the manifolds in the length-scale units of our code, and can therefore
be used to calibrate the sizes of the curvatures of the geometries.
\begin{table}[!hbt]
  \caption{Compact non-orientable manifolds included in this
    study. Also listed are the average scalar curvature $\langle
    \tilde R\,\rangle$ defined in Eq.~(\ref{e:AvarageRtildeDef}), the
    physical volumes $\mathcal{V}$ defined in
    Eq.~(\ref{e:PhysicalVolumeV}), and the cosmological constants
    $\Lambda$ determined by Eq.~(\ref{e:CMCParameterChoices})
    for the geometries with $K=-1$ constructed on each manifold.
    \label{t:TableI} }
  \begin{center}
  \begin{tabular}{lccc}
    Manifold & $\qquad\langle \tilde R\, \rangle\qquad$
    & $\qquad\mathcal{V}\qquad$  & $\qquad\Lambda\qquad$
 \\
 \hline
 \vspace{-7pt}\\
    $P^2\#\, P^2\times S^1$   &  $3.7\times 10^{-14}$ & \, 4.000  & 0.3333 \\
    $P^2\times S^1$           &  1.932             & \, 5.620  & 1.2992 \\
    $S^2\tilde\times S^1$     &  2.693             & \, 9.349  & 1.6797 \\
    $P^2\#\,T^2\times S^1$    & -3.866             & \, 4.113  & -1.5998 \\
 \hline
  \end{tabular}
  \end{center}
\end{table}

\begin{figure}[!h]
  \centering
  \includegraphics[height=0.32\textwidth]{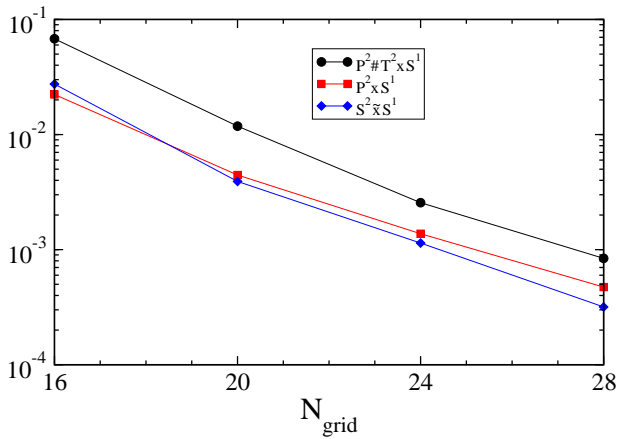}
\caption{\label{f:ConvergenceCCCons} This figure illustrates the norm
  of the Hamiltonian constraint, $||\,\mathcal{H}\,||$ defined in
  Eq.~(\ref{e:EinsteinConstraint}), as functions of the numerical
  resolution $N_\mathrm{grid}$ for the non-orientable manifolds
  $P^2\#\,T^2\times S^1$, $P^2\times S^1$, and $S^2\tilde\times S^1$.
  The values of $||\,\mathcal{H}\,||$ for the manifold
  $P^2\#\,P^2\times S^1$ are smaller than $6\times 10^{-10}$ for all
  values of $N_\mathrm{grid}$, and appears to be limited by double
  precision roundoff errors. For the other manifolds
  $||\,\mathcal{H}\,||/|\langle \tilde R\rangle|$ gives a
  dimensionless measure of constraint violations using the values of
  $\langle \tilde R\rangle $ (which are all of order unity) from
  Table~\ref{t:TableI}.}
\end{figure}

The accuracy of the resulting numerical solutions can be evaluated by
measuring how well they satisfy the constraints.  The Hamiltonian
constraint, $\mathcal{H}$, for the initial value problem considered
here is given by the expression
\begin{equation}
  \mathcal{H}=R +\textstyle\frac{2}{3}\left(K^2 - 3\Lambda\right) = R
  - \langle\tilde R\rangle,
\end{equation}
(see Ref.~\cite{Zhang2022}).  Consequently the Hamiltonian constraint
norm $||\,\mathcal{H}\,||$ defined by,
\begin{equation}
  ||\,\mathcal{H}\,||^2 = \mathcal{V}^{-1}\int \left(R - \langle\tilde R\,
  \rangle\right)^2 \!\sqrt{g}\, d^{\,3}x, 
  \label{e:EinsteinConstraint}
\end{equation}
is a useful tool for measuring the accuracy of the numerical
solutions.  Figure~\ref{f:ConvergenceCCCons} illustrates the values of
$||\,\mathcal{H}\,||$ as a function of the spatial resolution
$N_\mathrm{grid}$ (the number of grid points in each direction of each
multi-cube region) for each of the manifolds studied here.  These
numerical solutions are very time consuming for the higher resolution
cases, and it is possible that the final results reported here could
have been improved somewhat with more computer time or perhaps by
setting somewhat different parameters in the \verb!PETSc!
solvers~\cite{petsc-user-ref} used by the \verb!SpEC! code.
The results illustrated in Fig.~\ref{f:ConvergenceCCCons} show that
the numerical methods used in this study produce reasonably accurate,
numerically convergent solutions to the Einstein initial value
equations.%  

  The Hamiltonian constraint norm, $||\,\mathcal{H}\,||$, is
a measure of the homogeneity of the spatial Ricci scalar curvature $R$
for these solutions.  Therefore these initial data have good
homogeneity in their scalar Ricci curvatures, $R$, as well as exact
homogeneity in their extrinsic curvature traces, $K$.  The homogeneity of
these quantities are a necessary, but not sufficient, condition to
ensure that the spacetimes evolved from these initial data remain
homogeneous.  In a homogeneous spacetime the initial data for the
entire spatial Ricci tensor, $R_{\,ij}$, must also be homogeneous, not
just its trace, $R$.

The homogeneity of the full spatial Ricci tensor can be explored
further using the curvature scalar $\mathcal{R}$ defined by
\begin{equation}
  \mathcal{R}^2 = 3 R_{ij}R^{ij}.
  \label{e:CurvatureScalarDef}
\end{equation}
Its spatial average $\langle\mathcal{R}\,\rangle$ and spatial variations
$\Delta \mathcal{R}$ are defined by
\begin{eqnarray}
  \langle\mathcal{R}\,\rangle
  &=& \frac{1}{\mathcal{V}}\int \sqrt{g}\,\mathcal{R}\,d^{\,3}x,
  \label{e:CurvatureScalarAve}\\
  \Delta\mathcal{R} &=& \mathcal{R} - \langle\mathcal{R}\rangle.
  \label{e:CurvatureScaleFluct}
\end{eqnarray}
An $L_2$ norm of these variations, $||\,\Delta\mathcal{R}\,||$
is defined by,
\begin{equation}
  ||\,\Delta\mathcal{R}\,||^2
  = \frac{1}{\mathcal{V}}\int \sqrt{g}\left(\Delta\mathcal{R}\right)^2 d^{\,3}x,
  \label{e:CurvatureScaleFluctNorm}
\end{equation}
is a useful measure of the homogeneity of $\mathcal{R}$.
Table~\ref{t:TableII} compares the spatial variations in the Ricci
scalar curvature, $||\,\mathcal{H}\,||$, the average values of the
curvature scalar $\langle\mathcal{R}\,\rangle$, and the norm of the
variations of the curvature scalar $||\,\Delta \mathcal{R}\,||$ in the
initial data computed for each of the manifolds included in this
study.  These results show that while the full Ricci tensor in the
initial data for the $P^2\#\,P^2\times S^1$ manifold is homogeneous,
this is not the case for the other manifolds included in this study.
\begin{table}[!hbt]
  \caption{ This table lists the average spatial variations in the
    scalar curvature $||\,\mathcal{H}\,||$ defined in
    Eq.~(\ref{e:EinsteinConstraint}), the average curvature scalar
    $\langle\mathcal{R}\,\rangle$ defined in
    Eq.~(\ref{e:CurvatureScalarAve}), and the average spatial
    variations in the curvature scalar $||\,\Delta \mathcal{R}\,||$
    defined in Eq.~(\ref{e:CurvatureScaleFluctNorm}) for the
    $N_\mathrm{grid}=28$ initial data on the manifolds included in
    this study.
    \label{t:TableII} }
  \begin{center}
  \begin{tabular}{lccc}
    Manifold 
    & $\qquad||\,\mathcal{H}\,||\qquad$
    & $\qquad\langle\mathcal{R}\,\rangle\qquad$
    & $\quad||\,\Delta \mathcal{R}\,||\quad$  \\
 \hline
 \vspace{-7pt}\\
    $P^2\#\, P^2\times S^1$   & \, $6.5\times 10^{-11}$
         & $5.9\times 10^{-11}$ & $5.7\times 10^{-11}$\\
    $P^2\times S^1$           & \, $4.7\times 10^{-4}$ & $4.40$ & $3.63$\\
    $S^2\tilde\times S^1$     & \, $3.1\times 10^{-4}$ & $5.25$ & $2.82$\\
    $P^2\#\,T^2\times S^1$    & \, $8.5\times 10^{-4}$ & $9.37$ & $3.74$ \\
 \hline
  \end{tabular}
  \end{center}
\end{table}

\begin{figure}[h] 
  \centering
  \subfigure{ 
    \includegraphics[height=0.29\textwidth]{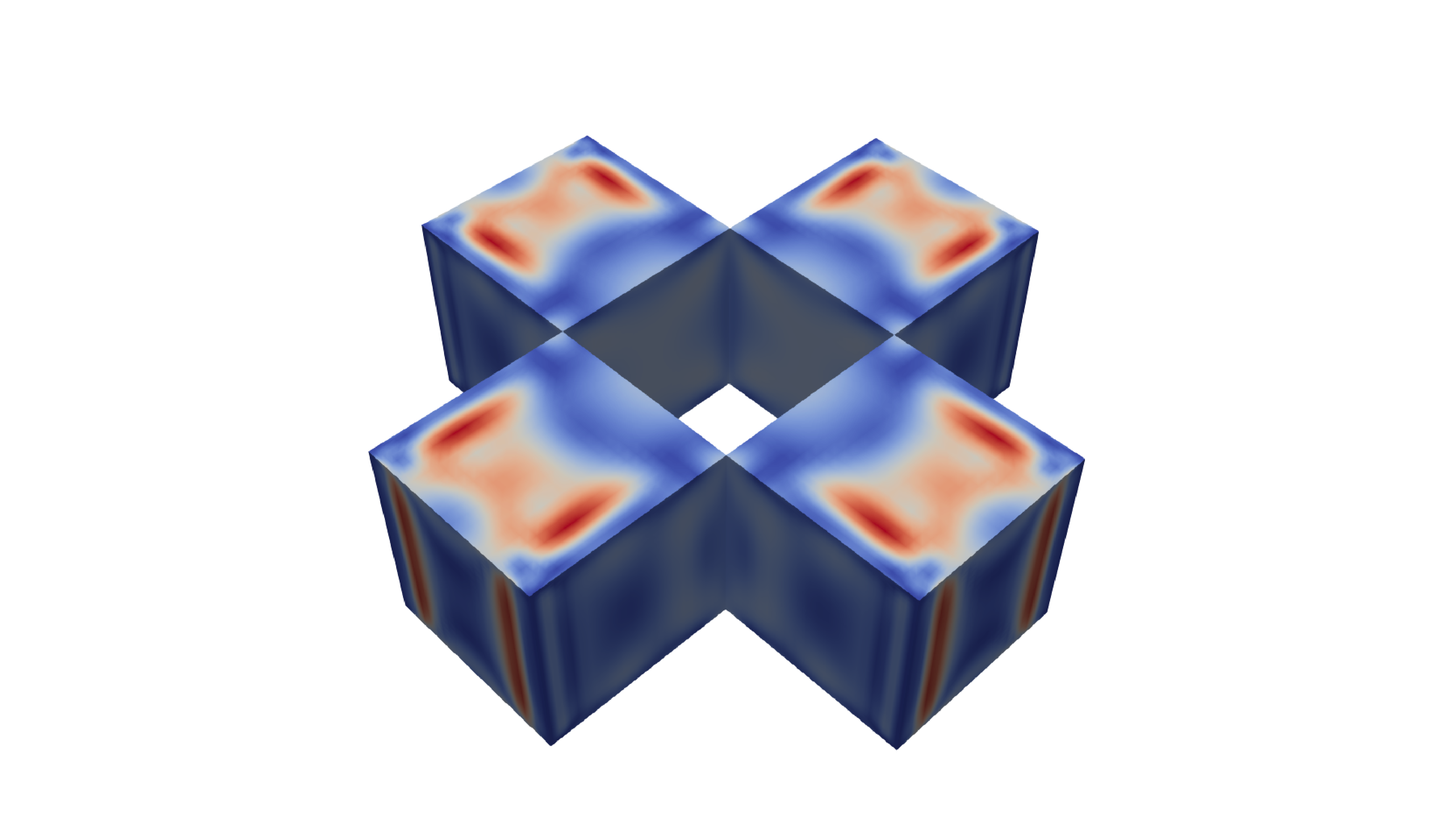}
  }
  \hspace{0.5cm}
  \caption{\label{f:P2xS1_DeltaEigenvalues} This figure illustrates
    the inhomogeneity of the variations in the curvature scalar
    $\Delta\mathcal{R}$ defined in Eq.~(\ref{e:CurvatureScaleFluct})
    for the initial data on the $P^2\#\,T^2\times S^1$ manifold.}
\end{figure}

The reason the spatial Ricci tensor $R_{\,ij}$ fails to be
homogeneous, even for the physical metrics constructed by solving the
Yamabe problem, can be traced to the properties of the conformal
metric $\tilde g_{ij}$.  The conformal metrics used for this study
were the reference metrics constructed to define the differentiable
structures on the multi-cube representations of the manifolds.  These
reference metrics were constructed from the basic structure of the
multi-cube representation.  They were designed to be $C^{2-}$ across
all the cube interfaces, but they were not designed to control the
spatial dependence of the curvature within each cube.  The conformal
transformation used to solve the Yamabe problem forced homogeneity on
the trace of the spatial Ricci tensor, but did not ensure the full
Ricci tensor was homogeneous in most cases.
Figure~\ref{f:P2xS1_DeltaEigenvalues} illustrates the initial values
of the curvature scalar variation, $\Delta\mathcal{ R}$, for the
$P^2\#\,T^2\times S^1$ manifold.  The imprint of the multi-cube
structure on the conformal metric $\tilde g_{ij}$ is clearly visible
in this figure.  The failure of the method used in this study to
construct simple homogeneous initial data prevents us from
constructing homogeneous cosmological models on most manifolds.

The curvature scalar $\mathcal{R}$ can also be used to detect
anisotropy in these geometries. It's value is the same as the
standard Ricci scalar curvature, $(\mathcal{R})^2=R^2$, if the space
is isotropic and all the eigenvalues of the Ricci tensor are the same.
A rough measure of the isotropy of these spaces can therefore be
obtained by comparing the average values of the Ricci scalar
curvatures $\langle\tilde R\,\rangle$ from Table~\ref{t:TableI} with
the average values of the curvature scalar
$\langle\mathcal{R}\,\rangle$ from Table~\ref{t:TableII}.  These
results show that the initial data on the $P^2\#\,P^2\times S^1$
manifold is isotropic as well as homogeneous, but this is not the case
for the other manifolds.  The $\mathcal{R}^2$ curvature scalar is
proportional to the sum of the squares of the eigenvalues of the Ricci
tensor.  In the $P^2\#\,P^2\times S^1$ manifold $\langle \mathcal{R}
\rangle= 0$ to roundoff error accuracy, therefore the geometry on this
manifold is flat.

%%%%%%%%%%%%%%%%%%%%%%%%%%%%%%%%%%%%%%%%%%%%%%%%%%%%%%%%%%%%%%%%%%%%%%%%%%%%%%%
% Non-Orientable Cosmological Models
%%%%%%%%%%%%%%%%%%%%%%%%%%%%%%%%%%%%%%%%%%%%%%%%%%%%%%%%%%%%%%%%%%%%%%%%%%%%%%%
\section{Simple Non-Orientable Cosmological Models}
\label{s:NonOrientableCosmologies}
%%%%%%%%%%%%%%%%%%%%%%%%%%%%%%%%%%%%%%%%%%%%%%%%%%%%%%%%%%%%%%%%%%%%%%%%%%%%%%%

This section describes the methods used to construct simple
non-orientable cosmological models by solving the Einstein evolution
equations using the initial data prepared in Sec.~\ref{s:InitialData}.
The spatial metrics, $g^0_{ij}$, produced by those solutions to the
Einstein constraint equations are used in two ways.  First, they are
used to construct the initial data for the spacetime metric
$\psi^0_{ab}$:
\begin{equation}
  ds^2=\psi^0_{ab}dx^adx^b = - dt^2 + g^0_{ij}dx^idx^j,
  \label{e:psi0Def}
\end{equation}
where the indices from the first part of the alphabet, e.g. $a$, $b$, $c$,
$...$, range over the four-dimensional spacetime coordinates, while the
indices from the later parts of the alphabet, e.g. $i$, $j$, $k$, $...$, range
over the three-dimensional spatial coordinates on each t = constant
hypersurface.  Second, the metric $\psi^0_{ab}$ is also used to define
the time-independent spacetime reference metric,
$\tilde\psi_{ab}\equiv\psi^0_{ab}$ that defines the differential
structure in these four-dimensional manifolds.
  
The representation of the Einstein equation used to perform these
evolutions is the covariant first-order symmetric-hyperbolic
representation developed in Ref.~\cite{Lindblom2014}.  The salient
features of this representation are summarized in
Appendix~\ref{s:SymmetricHyperbolicSystem}.  The dynamical fields in
this representation are the spacetime metric, $\psi_{ab}$, its time
derivative, $\Pi_{ab}=-t^c\tilde\nabla_c\psi_{ab}$, and its spatial
derivatives $\Phi_{iab}=\tilde\nabla_i\psi_{ab}$, where $t^c$ is the
future directed timelike unit normal ($t^a t^b \psi_{ab}=-1$) to the
$t =$ constant hypersurfaces.  The derivative, $\tilde\nabla_a$, used
in these definitions is the torsion-free covariant derivative
compatible with the reference metric:
$\tilde\nabla_c\tilde\psi_{ab}=0$.  The values of the dynamical fields
on the initial spacelike hypersurface used to start these evolutions
are given by,
\begin{eqnarray}
  \psi^0_{ab}&=& \tilde\psi_{ab},
  \label{e:Initialpsi}\\
  \Pi^0_{ta}&=&0,
  \label{e:InitialPita}\\
  \Pi^0_{ij}&=&\tfrac{1}{3}K^0 g^0_{ij},
  \label{e:InitialPiij}\\
  \Phi^0_{iab}&=&0,
  \label{e:InitialPhi}
\end{eqnarray}
where $K^0$ is the trace of the extrinsic curvature used to compute
the initial data in Sec.~\ref{s:InitialData}.

The first-order symmetric-hyperbolic representation of the Einstein
system has a large number of constraints, which are described in some
detail in Appendix~\ref{s:SymmetricHyperbolicSystem}.  The overall
magnitude of these constraints is summarized by a quantity
$\mathcal{C}_\psi$ defined in Eq.~(\ref{e:CpsiDef}).  The
dimensionless norm $||\,\mathcal{C}_\psi\,||$ defined in
Eq.~(\ref{e:CpsiNormDef}) provides a useful measure of the fractional
constraint violation errors in these cosmological models that is
monitored during each numerical evolution.

An important goal of this study is to compare the simple
non-orientable cosmological models constructed here with the standard
orientable homogeneous and isotropic models.  The spacetime metric for
the standard cosmological models can be written in the form,
\begin{equation}
  ds^2 = -d\eta^2 + a^2(\eta)L^2\left[d\chi^2 + \Xi^2(\chi)
    \left(d\theta^2+\sin^2\theta d\varphi^2\right)\right],
  \label{e:FriedmanMetric}
\end{equation}
where $L$ is a characteristic length scale, $a(\eta)$ is a
dimensionless scale factor, and $\Xi(\chi)$ is given by
\begin{equation}
  \Xi(\chi) = \left\{
  \begin{array}{ll}
    \sin\chi,\,\,&\mathrm{for}\,\, k=+1,\\
    \chi,\,\,&\mathrm{for}\,\, k=0,\\
    \sinh\chi,\,\,&\mathrm{for}\,\, k=-1,
  \end{array}\right.
\end{equation}
for the orientable manifolds with positive, zero, and negative spatial
curvatures respectively.  The Einstein equations for these geometries
are given by
\begin{eqnarray}
  \left(\frac{d a}{d\eta}\right)^{\!2} &=& -\frac{k}{L^2} + \frac{\Lambda}{3}\,a^2,
  \label{e:Friedman1}
  \\
  \frac{d^{\,2}a}{d\eta^{\,2}} &=& \frac{\Lambda}{3}\,a,
  \label{e:Friedman2}
\end{eqnarray}
in a spacetime where the cosmological constant $\Lambda$ dominates
over the matter content of the universe.

The properties of the non-orientable cosmological models resulting
from the numerical evolutions performed for this study are most easily
understood by expressing the resulting spacetime metric, $\psi_{ab}$,
in the three-plus-one ADM form~\cite{ADM1962}:
\begin{eqnarray}
  ds^2 &=& \psi_{ab}dx^adx^b\nonumber\\
  &=& -N^2dt^2 + g_{ij}(dx^i+N^idt)(dx^j+N^jdt),\qquad
  \label{e:ADMMetric}
\end{eqnarray}
where $N$ is the lapse, $N^i$ the shift, and $g_{ij}$ the spatial
metric on each $t =$ constant hypersurface.  There is no {\sl a
  priori} relationship between the $t=$ constant hypersurfaces
produced by the numerical evolutions, and the $\eta=$ constant
hypersurfaces used in the descriptions of the simple cosmological
models.  The $\eta=$ constant hypersurfaces are defined by the
homogeneity and isotrophy of the spatial geometries in those simple
models.  The cosmological models produced by the numerical evolutions
will only be comparable to the simple analytic models in those cases
where the $t=$ constant hypersurfaces also have homogeneous and
isotropic spatial geometries.

In order to facilitate the comparison between the non-orientable
cosmological solutions computed in this study and the simple
cosmological models it will be helpful to define a scale factor $a(t)$
that measures how the determinant of the spatial metric, $g(t)=\det
g_{ij}(t)$, evolves with time,
\begin{equation}
  a(t)^6=\frac{g(t)}{g(0)}.
  \label{e:ScaleFactorDef}
\end{equation}
The spatial average $\langle a(t)\rangle$, defined by
\begin{equation}
  \langle a(t)\rangle^3 = \frac{\int\sqrt{g(t)}\,d^{\,3}x}
          {\int\sqrt{g(0)}\,d^{\,3}x}
          = \frac{\mathcal{V}(t)}{\mathcal{V}(0)},
  \label{e:ScaleFactorAverage}
\end{equation}
is a generalization of the scale factor used in the standard
orientable homogeneous and isotropic cosmological models. 
The spatial variations in $a(t)$, as defined in
Eq.~(\ref{e:ScaleFactorDef}), can be measured with respect to its
spatial average, $\langle a(t)\rangle$, by evaluating the norm
$||\,\Delta a\,||$ defined by
\begin{equation}
  ||\,\Delta a\,||^2 = \frac{\int \left[a(t)-\langle a(t)\rangle\right]^2
  \sqrt{g}\,d^{\,3}x}{\langle a(t)\rangle^2\,\mathcal{V}(t)},
  \label{e:DeltaaDef}
\end{equation}
Similarly the homogeneity of the lapse $N$ can be evaluated:
\begin{eqnarray}
  ||\,\Delta N\,||^2 &=& \frac{\int \left[N(t) - \langle N(t)\rangle\right]^2
  \sqrt{g}\,d^{\,3}x}{\langle N(t)\rangle^2\,\mathcal{V}(t)},
  \label{e:DeltaNDef}
\end{eqnarray}
where $\langle N(t)\rangle$ is defined by,
\begin{eqnarray}
\langle N(t) \rangle &=& \frac{\int N(t)   
  \sqrt{g}\,d^{\,3}x}{\mathcal{V}(t)}.
  \label{e:NAverageDef}
\end{eqnarray}
The homogeneity of other properties of the spatial geometries like the
the trace of the extrinsic curvature, $K(t)$, the spatial Ricci
scalar, $R(t)$, and the curvature scalar $\mathcal{R}(t)$ defined in
Eq.~(\ref{e:CurvatureScalarDef}) can also be evaluated by monitoring
their spatial variations, defined in analogy with
Eqs.~(\ref{e:DeltaaDef}) and (\ref{e:DeltaNDef}).

For the homogeneous non-orientable cosmological models constructed in
this study, it will be interesting to compare the numerical evolution
of $\langle a(t)\rangle$ with the analogous solutions for $a(\eta)$
from Eqs.~(\ref{e:Friedman1}) and (\ref{e:Friedman2}).  These
comparisons are complicated by the fact that the gauge choice made to
ensure stable numerical evolutions do not keep the lapse $N(t)$ fixed
at its initial value.  Consequently the relationship between the
numerical time coordinate, $t$, and the time coordinate, $\eta$, used
on the standard cosmological models must be determined before
comparisons between the homogeneous cosmological models can be made.

%%%%%%%%%%%%%%%%%%%%%%%%%%%%%%%%%%%%%%%%%%%%%%%%%%%%%%%%%%%%%%%%%%%%%%%%%%%%%%%
\subsection{Simple Homogeneous Cosmological Model}
\label{s:HomogeneousModels}
%%%%%%%%%%%%%%%%%%%%%%%%%%%%%%%%%%%%%%%%%%%%%%%%%%%%%%%%%%%%%%%%%%%%%%%%%%%%%%%
%Numerical Results for P2+P2xS1 evolutions
%%%%%%%%%%%%%%%%%%%%%%%%%%%%%%%%%%%%%%%%%%%%%%%%%%%%%%%%%%%%%%%%%%%%%%%%%%%%%%%

\begin{figure}[!h]
  \centering
  \subfigure{
    \includegraphics[height=0.32\textwidth]{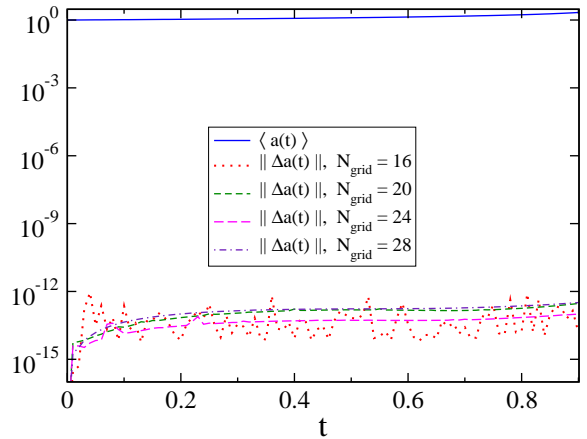}
  }
  \caption{\label{f:P2+P2xS1_ScaleFactor} This graph illustrates the
    average scale factor, $\langle a(t)\rangle$, and its spatial
    variations, $||\,\Delta a(t)\,||$, defined in
    Eqs.~(\ref{e:ScaleFactorAverage}) and (\ref{e:DeltaaDef}) for the
    evolutions on the $P^2\#\, P^2\times S^1$ manifold.  The results
    of these evolutions for the spatial averages, $\langle
    a(t)\rangle$ using different resolutions are indistinguishable on
    the scale of this graph, so only the result from the highest
    resolution is shown.}
\end{figure}

The only homogeneous and isotropic initial data constructed in
Sec.~\ref{s:InitialData} were the data for the $P^2\#\,P^2\times S^1$
manifold.  These initial data were evolved numerically until the
spatial volume of the space grew to ten times its initial value.
Figure~\ref{f:P2+P2xS1_ScaleFactor} illustrates this volume increase
by showing that the scale factor $a(t)$ increases from its initial
value $a(0)=1$ to the value $a(0.9)\approx2.154$ at the end of the
evolution.  These evolutions were performed using multi-cube grids
with resolutions $N_\mathrm{grid}=\{16,20,24,28\}$ in each spatial
direction.

\begin{figure}[!h]
  \centering \subfigure{
    \includegraphics[height=0.32\textwidth]{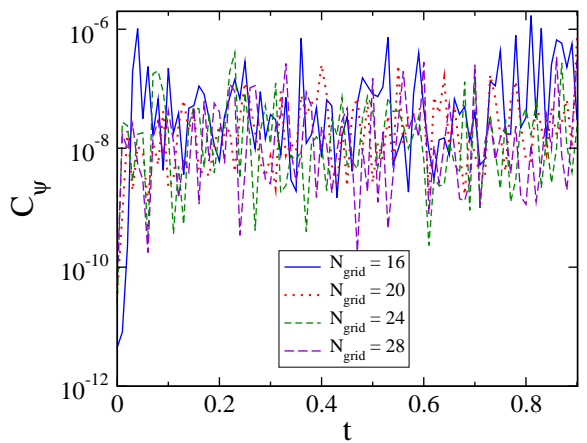}
  }
  \caption{\label{f:P2+P2xS1_GhCeCov} This graph
    illustrates the constraint norm, $||\mathcal{C}_\psi||$, defined in
    Eq.~(\ref{e:CpsiNormDef}) for the evolutions on the $P^2\#\,P^2\times
    S^1$ manifold.     }
\end{figure}

Figure~\ref{f:P2+P2xS1_GhCeCov} illustrates the constraint norms
$||\,\mathcal{C}_\psi||$ for these evolutions.  These norms are all at
about the $10^{-8}$ to $10^{-7}$ level and do not show significant (if
any) convergence as $N_\mathrm{grid}$ increases.  The lower limit on
these norms is probably being set by the $10^{-7}$ time stepping
tolerance used in these evolutions.

\begin{figure}[t]
  \centering
  \subfigure{
    \includegraphics[height=0.32\textwidth]{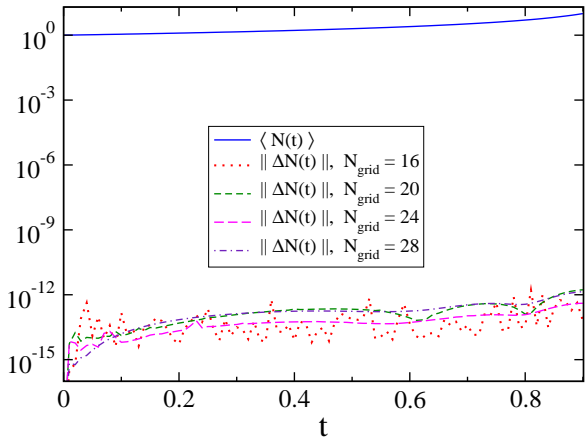}
  }
  \caption{\label{f:P2+P2xS1_Lapse} This graph illustrates the average
    value of the lapse, $\langle N(t)\rangle$, and its spatial
    variations, $||\,\Delta N(t)\,||$, defined in
    Eqs.~(\ref{e:NAverageDef}) and (\ref{e:DeltaNDef}).  }
\end{figure}

Figures~\ref{f:P2+P2xS1_ScaleFactor} and \ref{f:P2+P2xS1_Lapse}
illustrate the evolutions of the average scale factor $\langle
a(t)\rangle$, the average lapse $\langle N(t) \rangle$ and their
spatial variations $||\Delta a(t)||$ and $||\Delta N(t)||$ for the
evolutions on the $P^2\#\,P^2\times S^1$ manifold.  These graphs
clearly show that both the scale factor $a(t)$ and the lapse $N(t)$
remain homogeneous at the double precision roundoff error level
throughout these evolutions.  Similar levels of homogeneity were also
maintained (but not displayed to save space) by the other geometric
scalars in these solutions, i.e. the trace of the extrinsic curvature
$K(t)$, the trace of the spatial Ricci curvature $R(t)$, and the
curvature scalar $\mathcal{R}(t)$.  The norm of the shift $N^i(t)$ was
also found to be less than $4.5\times 10^{-13}$ throughout these
evolutions.

\begin{figure}[h]
  \centering
  \subfigure{
    \includegraphics[height=0.28\textwidth]{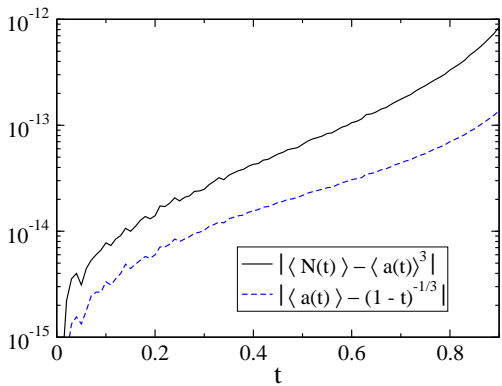}
  }
  \caption{\label{f:P2+P2xS1_Analytic} The solid (black) curve in this
    graph demonstrates that $|\langle N\rangle - \langle
    a\rangle^3|<9\times 10^{-13}$ for the $N_\mathrm{grid}=28$
    evolution on the $P^2\#\,P^2\times S^1$ manifold.  Similarly the
    dashed (blue) curve in this graph demonstrates that $|\,\langle
    a(t) \rangle - (1-t)^{-1/3}\,|< 2\times 10^{-13}$ for this
    evolution.}
\end{figure}

Figure~\ref{f:P2+P2xS1_Lapse} shows that the lapse $N(t)$ does
not remain constant during the evolutions, increasing from its initial
value $N(0)=1$ to the value $N(0.9)=10.0$ at the end.  This implies that
the time coordinate $t$ used in the numerical evolutions is not the
same as the time coordinate $\eta$ used in the standard
representations of the homogeneous cosmological models. Empirically we
find that $\langle N \rangle \approx \langle a\rangle^3$ for the
evolutions on the $P^2\#\,P^2\times S^1$ manifold, as demonstrated in
Fig.~\ref{f:P2+P2xS1_Analytic} for the $N_\mathrm{grid}=28$ evolution.

The scale factor $a(\eta)$ is determined in the standard homogeneous
cosmological models by Eq.~(\ref{e:Friedman1}).  The constant $k$ in
that equation vanishes, $k=0$, for the flat geometry on this
$P^2\#\,P^2\times S^1$ manifold (see Table~\ref{t:TableI}), thus
Eq.~(\ref{e:Friedman1}) simplifies in this case to
\begin{equation}
  \frac{da}{d\eta} = \sqrt{\frac{\Lambda}{3}}\,a.
  \label{e:flatFriedman}
\end{equation}
Since $\langle N \rangle = \langle a\rangle^3$ for these evolutions
(which are essentially homogeneous as shown in
Figs.~\ref{f:P2+P2xS1_ScaleFactor} and \ref{f:P2+P2xS1_Lapse}) it
follows that $d\eta/dt = a^3$.  Therefore Eq.~(\ref{e:flatFriedman})
can be transformed into an equation for $a(t)$:
\begin{equation}
  \frac{da}{dt} = \sqrt{\frac{\Lambda}{3}}\,a^4.
  \label{e:flatFriedman2}
\end{equation}
It is straightforward to integrate this equation analytically, with
the result
\begin{equation}
  a(t)=\left(1-3\sqrt{\frac{\Lambda}{3}}\, t\right)^{-1/3}.
  \label{e:flatFriedman_analytic}
\end{equation}
The value of the cosmological constant $\Lambda$ used for these
evolutions is $\Lambda=1/3$, see Table~\ref{t:TableI}.  Therefore
the analytic expression for $a(t)$ from Eq.~(\ref{e:flatFriedman_analytic})
reduces to
\begin{equation}
  a(t)=\left(1- t\right)^{-1/3}.
  \label{e:flatFriedman_analytic2}
\end{equation}
Figure~\ref{f:P2+P2xS1_Analytic} shows that the numerical $\langle a(t)\rangle$
for these evolutions agrees with this analytical expression to within
$2\times 10^{-13}$ for the $N_\mathrm{grid}=28$ evolution.

%%%%%%%%%%%%%%%%%%%%%%%%%%%%%%%%%%%%%%%%%%%%%%%%%%%%%%%%%%%%%%%%%%%%%%%%%%%%%%%
\subsection{Simple Inhomogeneous Cosmological Models}
\label{s:InhomogeneousModels}
%%%%%%%%%%%%%%%%%%%%%%%%%%%%%%%%%%%%%%%%%%%%%%%%%%%%%%%%%%%%%%%%%%%%%%%%%%%%%%%
%Simple Non-Hmogeneous Cosmological Models
%%%%%%%%%%%%%%%%%%%%%%%%%%%%%%%%%%%%%%%%%%%%%%%%%%%%%%%%%%%%%%%%%%%%%%%%%%%%%%%

The initial data constructed in Sec.~\ref{s:InitialData} on the
$P^2\times S^1$, $S^2\tilde\times S^1$, and $P^2\#\,T^2\times S^1$
manifolds are significantly inhomogeneous, so the cosmological models
evolved from them are not at all similar to the structure of our
universe.  These inhomogeneous models do, however, provide useful
tests of the numerical methods that have been developed to solve
Einstein's equation numerically on manifolds with arbitrary spatial
topologies.  Those methods had not previously been tested on
significantly non-linear evolutions like the inhomogeneous
cosmological models constructed here.

The initial data on the $P^2\times S^1$ and $S^2\tilde\times S^1$
manifolds were evolved in these tests until their spatial volumes
increased by a factor of ten.  The spatial volumes in the evolutions
on the $P^2\#\,T^2\times S^1$ manifold increase initially, but then
reach a maxima when their volumes are about 1.28 times their initial
values.  The volumes in the $P^2\#\,T^2\times S^1$ evolutions then
contract and their evolutions were continued until their volumes
returned to their initial values.

Perhaps the most interesting features of these inhomogeneous
evolutions are the constraint norms $||\,\mathcal{C}_\psi\,||$.  The
evolution of these norms are shown in Fig.~\ref{f:S2xtildeS1_GhCeCov}
for the $S^2\tilde\times S^1$ manifold.  The results for the
$S^2\tilde\times S^1$ and $P^2\#\,T^2\times S^1$ manifolds are
qualitatively similar, and therefore will not be displayed to conserve
space.  The most important feature of these constraint norm graphs is
the fact that they show excellent numerical convergence.  These graphs
show that value of $||\,\mathcal{C}_\psi\,||$ is reduced by a factor
of about 0.8 as the spatial resolution $N_\mathrm{grid}$ is increased
by four in each successive higher resolution evolution.  This is the key
indicator of exponential convergence, and shows that the numerical
methods used in this study are producing exponentially convergent
solutions to the Einstein evolution equations.
\begin{figure}[h]
  \centering
  \subfigure{
    \includegraphics[height=0.32\textwidth]{./Fig9.eps}
  }
  \caption{\label{f:S2xtildeS1_GhCeCov} This graph illustrates the
    dimensionless constraint norm, $||\,\mathcal{C}_\psi\,||$, defined
    in Eq.~(\ref{e:CpsiNormDef}) for the evolutions on the
    $S^2\tilde\times S^1$ manifold.}
\end{figure} 
\begin{figure}[h]
  \centering
  \subfigure{
    \includegraphics[height=0.32\textwidth]{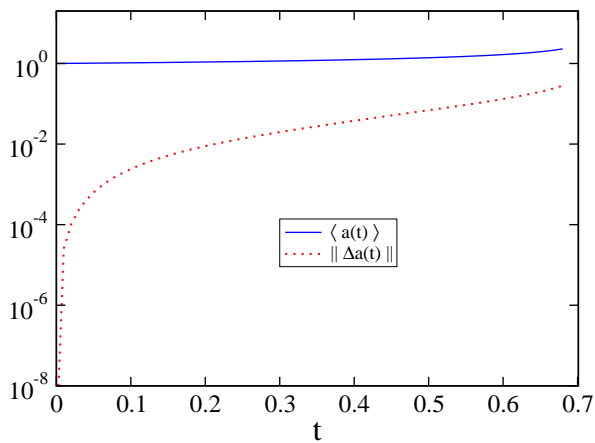}
  }
  \caption{\label{f:P2xS1_ScaleFactor} This graph illustrates the
    average scale factor, $\langle a(t)\rangle$, and its spatial
    variations, $||\,\Delta a(t)\,||$, defined in
    Eqs.~(\ref{e:ScaleFactorAverage}) and (\ref{e:DeltaaDef}) for the
    evolutions on the $P^2\times S^1$ manifold.}
\end{figure} 
Figure~\ref{f:P2xS1_ScaleFactor} illustrates the average scale factor
$\langle a(t)\rangle$ and the norm of its spatial variations
$||\,\Delta a(t)\,||$ defined in Eqs.~(\ref{e:ScaleFactorAverage}) and
(\ref{e:DeltaaDef}) for evolutions on the $P^2\times S^1$ manifold.
The results for the evolutions on the $S^2\tilde\times S^1$ and
$P^2\#\,T^2\times S^1$ manifolds are qualitatively similar. The
results of these evolutions using different resolutions are
indistinguishable on the scale of this graph, so only the results for
the $N_\mathrm{grid}=28$ case are shown.  The spatial variation in the
scale factor $||\,\Delta a(t)\,||$ increases from zero in the initial
data to about $0.28$ at the end of the evolution.  This illustrates
the significant inhomogeneity of these models.

\begin{figure}[h]
  \centering
  \subfigure{
    \includegraphics[height=0.32\textwidth]{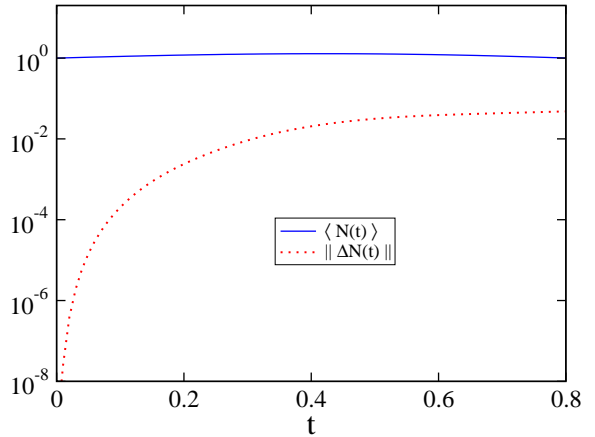}
 }
  \caption{\label{f:P2+T2xS1_Lapse} This graph illustrates the average
    value of the lapse, $\langle N(t)\rangle$, and its spatial
    variations, $||\,\Delta N(t)\,||$, defined in
    Eqs.~(\ref{e:DeltaNDef}) and (\ref{e:NAverageDef}) for the
    evolutions on the $P^2\# T^2\times S^1$ manifold. }
\end{figure}

Figure~\ref{f:P2+T2xS1_Lapse} illustrates the average value of the
lapse $\langle N(t)\rangle$ and the norm of its spatial variations
$||\,\Delta N(t)\,||$ defined in Eqs.~(\ref{e:DeltaNDef}) and
(\ref{e:NAverageDef}) for evolutions on the $P^2\#\,T^2\times S^1$
manifold.  The results of these evolutions using different resolutions
are indistinguishable on the scale of this graph, so only the results
for the $N_\mathrm{grid}=28$ case are shown.  The spatial variation in
the lapse $||\,\Delta N(t)\,||$ increases from zero in the initial
data to about 0.046, and the norm of the shift $N^i$ increases to
about $0.094$ by the ends of these evolutions.  The results for the
evolutions on the $P^2\times S^1$ and $S^2\tilde\times S^1$ manifolds
are qualitatively similar.

%

%%%%%%%%%%%%%%%%%%%%%%%%%%%%%%%%%%%%%%%%%%%%%%%%%%%%%%%%%%%%%%%%%%%%%%%%%%%%%%%
\section{Discussion}
\label{s:Discussion}
%%%%%%%%%%%%%%%%%%%%%%%%%%%%%%%%%%%%%%%%%%%%%%%%%%%%%%%%%%%%%%%%%%%%%%%%%%%%%%%

This study focused on applying and evaluating the numerical methods
developed in recent years to solve Einstein's equation on manifolds
with non-trivial
topologies~\cite{Lindblom2013,Lindblom2014,Lindblom2015,Zhang2022} by
applying them to a study of simple cosmological models on manifolds
with non-orientable spatial slices.  Solutions to Einstein's equation
were constructed on manifolds with spatial slices having the four
non-orientable topologies: : $P^2\times S^1$, $P^2\#P^2\times S^1$,
$P^2\# T^2\times S^1$, and $S^2\tilde\times S^1$.  The Einstein
constraint equations were solved numerically on these manifolds to
produce initial data with homogeneous spatial Ricci scalar curvatures
and constant extrinsic curvature traces.  These initial data were then
evolved numerically long enough to allow the spatial volumes of the
manifolds to increase by a factor of ten.  These numerical solutions
were shown to be exponentially convergent on the manifolds $P^2\times
S^1$, $P^2\# T^2\times S^1$, and $S^2\tilde\times S^1$.  The extremely
small constraint violations in the solutions on the $P^2\#P^2\times
S^1$ manifold appear to be limited by double precision roundoff
errors for the elliptic initial data solves, and time-stepping errors
for the hyperbolic evolutions.  The numerical solutions on the
manifold with spatial topology $P^2\#P^2\times S^1$ were shown to
produce homogeneous cosmological solutions that agree with high
precision locally to the corresponding Friedman cosmological model on
an orientable manifold.  The numerical solutions on the manifolds
$P^2\times S^1$, $P^2\# T^2\times S^1$, and $S^2\tilde\times S^1$ were
all shown to have significant inhomogeneities in their initial data,
and these inhomogeneities were shown to grow as those initial data
were evolved.  While these inhomogeneous solutions do not make good
cosmological models of our universe, they do serve as good tests of
our numerical methods and their implementation in the \verb!SpEC!
code.

This study also reveals serious limitations in the method used to
construct homogeneous initial data for cosmological models on
manifolds with non-trivial topologies.  In general, the reference
metrics constructed by the method described in
Ref.~\cite{Lindblom2015,Lindblom2022} are not well suited to serve as
the conformal metrics used to solve the initial data problem.  These
reference metrics have very inhomogeneous geometries that can not be
made homogeneous by a simple conformal transformation.  Better methods
must be developed for constructing suitable conformal metrics on these
manifolds.  At this point perhaps the most promising method would be
to smooth the current reference metrics by evolving them using Ricci
flow~\cite{Chow2004}.  We intend to explore this possibility in a
future study.

%%%%%%%%%%%%%%%%%%%%%%%%%%%%%%%%%%%%%%%%%%%%%%%%%%%%%%%%%%%%%%%%%%%%%%%%%%%%%%%
\appendix
%%%%%%%%%%%%%%%%%%%%%%%%%%%%%%%%%%%%%%%%%%%%%%%%%%%%%%%%%%%%%%%%%%%%%%%%%%%%%%%
\section{Covariant First-Order Symmetric-Hyperbolic Einstein System}
\label{s:SymmetricHyperbolicSystem}
%%%%%%%%%%%%%%%%%%%%%%%%%%%%%%%%%%%%%%%%%%%%%%%%%%%%%%%%%%%%%%%%%%%%%%%%%%%%%%%
\renewcommand{\theequation}{A.\arabic{equation}} % Redefines equation numbering
\setcounter{equation}{0} % Resets the equation counter for the appendix

This study uses a covariant representation of the Einstein equations
that enforces generalized harmonic gauge
conditions~\cite{Lindblom2014}.  The covariance of this representation
is essential for the study of solutions on manifolds with
multiple multi-cube coordinate patches, such as the non-orientable
manifolds studied here.  All the dynamical fields in this
representation are tensors, so they can be transformed in a
straightforward way across the interfaces between coordinate patches.
The dynamical fields, $u^\alpha$, for the first-order
symmetric-hyperbolic representation of the Einstein equations used in
this study are the collection of tensor fields,
\begin{equation}
  u^\alpha = \left\{ \psi_{ab}, \Pi_{ab}, \Phi_{iab}\right\},
  \label{e:UalphaDef}
\end{equation}
were $\psi_{ab}$ is the spacetime metric, and the tensors $\Pi_{ab}$
and $\Phi_{iab}$ represent its first derivatives,
\begin{eqnarray}
  \Pi_{ab} &=& -t^c\tilde\nabla_c\psi_{ab},
  \label{e:PiDef}\\
  \Phi_{iab}&=& \tilde\nabla_i\psi_{ab}. 
  \label{e:PhiDef}
\end{eqnarray} 
The derivative $\tilde\nabla_c$ used in these expressions is the
torsion-free covariant derivative compatible with the reference metric
$\tilde\psi_{ab}$: $\tilde\nabla_c\tilde\psi_{ab}=0$.  This reference
metric is also used to define the Jacobians used to transform tensor
fields across the boundary interfaces between multi-cube regions.  The
vector $t^c$ represents the unit timelike normal ($t^a t^b
\psi_{ab}=-1$) to the spacelike hypersurfaces used to perform the
evolutions.  The lower range of Latin indices, e.g. $a$, $b$, $c$,
..., range over the four spacetime coordinates, while the higher range
of indices, e.g. $i$, $j$, $k$, ..., range over the three spatial
coordinates on each $t=$ constant hypersurface.  The Greek indices,
e.g. $\alpha$, $\beta$, $\gamma$, range over the fifty components of
the dynamical tensor fields $u^\alpha$.  The Einstein equations for
these fields can be written in the first-order form
\begin{equation}
  \partial_t u^\alpha
  + A^{k\alpha}{}_\beta({\bf u})\tilde\nabla_ku^\beta
  =F^\alpha({\bf u},{\bf \tilde\Gamma}, {\bf \partial\tilde\Gamma}),
  \label{e:EvolutionEqs}
\end{equation}
where the tensors $ A^{k\alpha}{}_\beta({\bf u})$ and $F^\alpha({\bf
  u},{\bf \tilde\Gamma}, {\bf \partial\tilde\Gamma})$ depend in
complicated ways on the dynamical fields $u^\alpha$, the reference
connection $\tilde\Gamma^c{}_{ab}$ and its derivatives
$\partial_d\tilde\Gamma^c{}_{ab}$.  The detailed expressions for this
representation of the covariant Einstein evolution system are given in
Ref.~\cite{Lindblom2014}.

The spacetime coordinates are determined in the generalized harmonic
representations of Einstein's equation by imposing a condition on the
difference between the connection $\tilde\Gamma^c{}_{ab}$ associated
with the reference metric $\tilde \psi_{ab}$ and the physical
connection $\Gamma^c{}_{ab}$ associated with the physical spacetime
metric $\psi_{ab}$.  In particular the difference between these
connections is fixed by a gauge source function $H_a$:
\begin{equation}
H_a =  -\psi_{ad}\,\psi^{bc}\left(\Gamma^d{}_{bc}-\tilde\Gamma^d{}_{bc}\right).
  \label{e:HarmonicGageDef}
\end{equation}
(Note that the difference between any two connections is a tensor.)
In general $H_a$ may be any function that depends on the physical
metric $\psi_{ab}$ (but not its derivatives) and the reference metric
$\tilde\psi_{ab}$.  The gauge condition used for the numerical
evolutions performed in this study is the simple harmonic condition,
$H_a=0$.

The generalized harmonic evolution system contains a number of
constraints.  In particular the gauge condition,
Eq.~(\ref{e:HarmonicGageDef}) is in effect a constraint
\begin{equation}
  \mathcal{C}_a =H_a +
  \psi_{ad}\psi^{bc}\left(\Gamma^d{}_{bc}-\tilde\Gamma^d{}_{bc}\right).
  \label{e:C1Def}
\end{equation}
on the dynamical fields.  In addition, the equation that defines
$\Phi_{iab}$, Eq. (\ref{e:PhiDef}), is also a constraint
\begin{equation}
  \mathcal{C}_{iab} = \tilde\nabla_i\psi_{ab}-\Phi_{iab}.
  \label{e:C3Def}
\end{equation}
These primary constraints, $\mathcal{C}_a$ and $\mathcal{C}_{iab}$,
satisfy a second-order system of evolution equations as a consequence
of the Einstein evolution system, Eq.~(\ref{e:EvolutionEqs}). (See
Ref.~\cite{Lindblom2014} for details.)  This second-order constraint
evolution system can be converted to a first-order symmetric
hyperbolic system by introducing the secondary constraints,
\begin{eqnarray}
  \mathcal{F}_a &=& t^c\nabla_c\mathcal{C}_a,
  \label{e:F1Def}\\
  \mathcal{C}_{ia} &=& \nabla_i\mathcal{C}_a,
  \label{e:C2Def}\\
  \mathcal{C}_{ijab}&=&2\tilde\nabla_{[i}\mathcal{C}_{j]ab}.
  \label{e:C4Def}
\end{eqnarray}
The symmetric-hyperbolic evolution equation for the collection of
constraints,
\begin{equation}
  \mathcal{C}^\alpha=\left\{\mathcal{C}_a,\mathcal{C}_{iab},
  \mathcal{F}_a,\mathcal{C}_{ia},\mathcal{C}_{ijab}\right\},
  \label{e:ConstraintTotal}
\end{equation}
ensures that solutions to the Einstein Eq.~(\ref{e:EvolutionEqs}) that
satisfy the constraints, $\mathcal{C}^\alpha$, on an initial surface
will satisfy them throughout the evolution of those initial data.
Numerical solutions to the equations will of course contain (hopefully
small) violations of these constraints.  This study monitors the size
of these constraint violations by evaluating a norm constructed from
the composite constraint, $\mathcal{C}_\psi$, defined by,
\begin{eqnarray}
  \mathcal{C}_\psi^2 &=& \delta^{ab}\left(\mathcal{C}_a\mathcal{C}_b
  +\mathcal{F}_a\mathcal{F}_b\right)\nonumber\\
&&  +\tilde g^{ij}\delta^{ab}\delta^{cd}\left(\mathcal{C}_{iac}\mathcal{C}_{jbd}
    +\tfrac{1}{4}\tilde g^{kl}\mathcal{C}_{ikac}\mathcal{C}_{jlbd}\right).
    \qquad
  \label{e:CpsiDef}
\end{eqnarray}
This study evaluates a norm of $\mathcal{C}_\psi$,
$||\,\mathcal{C}_\psi\,||$, defined by
\begin{equation}
  ||\,\mathcal{C}_\psi\,||^2 = \frac{\int \mathcal{C}_\psi^2\sqrt{g}\,d^{\,3}x}
  {\int\mathcal{N}_\psi^2\sqrt{g}\,d^{\,3}x}.
  \label{e:CpsiNormDef}
\end{equation}
The quantity that appears in the denominator, $\mathcal{N}_\psi$,
is defined by
\begin{eqnarray}
  \mathcal{N}_\psi^2&=& \delta^{ab}\delta^{cd}\tilde g^{ij}\left(
  \partial_i\psi_{ac}\partial_j\psi_{bd}
  +  \partial_i\Pi_{ac}\partial_j\Pi_{bd}\right)\nonumber\\
  &&+\delta^{ab}\delta^{cd}\tilde g^{ij}
  \tilde g^{kl}\partial_i\Phi_{kac}\partial_j\Phi_{lbd}.
  \label{e:NpsiDef}
\end{eqnarray}
It has been included in the definition of $||\,\mathcal{C}_\psi\,||$ to 
provide a dimensionless measure of the fractional errors due to
constraint violations in the numerical solutions to the Einstein
evolution system.

%%%%%%%%%%%%%%%%%%%%%%%%%%%%%%%%%%%%%%%%%%%%%%%%%%%%%%%%%%%%%%%%%%%%%%%%%%%%%%%
\begin{acknowledgements}
  We thank Michael Holst for providing the computational resources
  used to perform the numerical calculations reported in this study.
  We also thank James Nester for several helpful conversations about
  the geometry of non-orientable manifolds.  L.L. was supported in
  part by grant No. 2407545 from the National Science Foundation to
  the University of California at San Diego, USA. F.Z. was supported
  by the National Key Research and Development Program of China grant
  2023YFC2205801, and the National Natural Science Foundation of China
  grants 12433001 and 12021003.
\end{acknowledgements}
%%%%%%%%%%%%%%%%%%%%%%%%%%%%%%%%%%%%%%%%%%%%%%%%%%%%%%%%%%%%%%%%%%%%%%%%%%%%%%%

\vfill\break

%%%%%%%%%%%%%%%%%%%%%%%%%%%%%%%%%%%%%%%%%%%%%%%%%%%%%%%%%%%%%%%%%%%%%%%%%%%%%%%
% References
%%%%%%%%%%%%%%%%%%%%%%%%%%%%%%%%%%%%%%%%%%%%%%%%%%%%%%%%%%%%%%%%%%%%%%%%%%%%%%%
\bibliographystyle{spphys}
\bibliography{../References/References}

%%%%%%%%%%%%%%%%%%%%%%%%%%%%%%%%%%%%%%%%%%%%%%%%%%%%%%%%%%%%%%%%%%%%%%%%%%%%%%%
\end{document}